\documentstyle[12pt]{article}
\newcommand{\extraspace}{\addtolength{\abovedisplayskip}{2mm}
                        \addtolength{\belowdisplayskip}{2mm}
                        \addtolength{\abovedisplayshortskip}{2mm}
                        \addtolength{\belowdisplayshortskip}{2mm}}
\newcommand{\be}{\begin{equation}\extraspace}

\newcommand{\ee}{\end{equation}}

\newcommand{\bea}{\begin{eqnarray}\extraspace}
\newcommand{\beastar}{\begin{eqnarray*}\extraspace}
\newcommand{\eea}{\end{eqnarray}}
\newcommand{\eeastar}{\end{eqnarray*}}

\newcommand{\nonu}{\nonumber \\[2mm]}

\newcommand{\stru}{\rule[0mm]{0mm}{6mm}}

\newcommand{\half}{\frac{1}{2}}
\newcommand{\thalf}{{\textstyle \frac{1}{2}}}
\newcommand{\tquart}{{\textstyle \frac{1}{4}}}

\newcommand{\threehalves}{{\textstyle \frac{3}{2}}}

\newcommand{\del}{\partial}
\newcommand{\eg}{{\it e.g.}}
\newcommand{\ie}{{\it i.e.}}

\newcommand{\vac}{|0\rangle}
\newcommand{\pp}{+ \!\!\! +}
\newcommand{\mm}{=}
\newcommand{\qq}{{q \over 2}}

\newcommand{\newsection}[1]{
\vspace{12mm}
\pagebreak[3]
\addtocounter{section}{1}
\setcounter{equation}{0}
\setcounter{subsection}{0}
\setcounter{footnote}{0}
\begin{flushleft}
{\large\bf \thesection. #1}
\end{flushleft}
\nopagebreak
\smallskip
\vspace{2mm}
\nopagebreak}

\newcommand{\newsubsection}[1]{
\vspace{8mm}
\pagebreak[3]

\addtocounter{subsection}{1}
\addcontentsline{toc}{subsection}{\protect
\numberline{\arabic{section}.\arabic{subsection}}{#1}}
\noindent{\sc \thesubsection. #1}
\nopagebreak
\vspace{6mm}
\nopagebreak}

\def\ZZ{Z\!\!\!Z} 		
\newcommand{\ts}{\textstyle}

\begin{document}
\baselineskip=17pt

\hfill {USC-94/21, UCSB-TH-94, PUPT-1523}

\hfill {hep-th/9412199}

\vskip .8cm
\begin{center}

{\Large Affine and Yangian Symmetries in}

\vskip 4mm

{\Large $SU(2)_1$ Conformal Field Theory%
\footnote{Lectures given at the 1994 Trieste Summer
School on High Energy Physics and Cosmology,
Trieste, July 1994.}}

\vskip .8 cm

{\large Peter Bouwknegt}

\vskip .3cm

{\sl Department of Physics and Astronomy,
     University of Southern California \\
     Los Angeles, CA 90089-0484}

\vskip .6 cm

{\large Andreas W.W. Ludwig}

\vskip .3cm

{\sl Department of Physics,
     University of California at Santa Barbara \\
     Santa Barbara, CA 93106}

\vskip .6 cm

{\large Kareljan Schoutens}

\vskip .3cm

{\sl Joseph Henry Laboratories, Princeton University \\
     Princeton, NJ 08544}

\vskip .7cm

{\bf Abstract}

\end{center}

\baselineskip=15pt

{\small
In these lectures, we study and compare two different
formulations of $SU(2)$, level $k=1$, Wess-Zumino-Witten
conformal field theory. The first, conventional, formulation
employs the affine symmetry of the model; in this
approach correlation functions are derived from the
so-called Knizhnik-Zamolodchikov equations.
The second formulation is based on an entirely different
algebraic structure, the so-called Yangian $Y(sl_2)$.
In this approach, the Hilbert space of the theory is
obtained by repeated application of modes of the
so-called spinon field, which has $SU(2)$ spin $j=\thalf$
and obeys fractional (semionic) statistics. We show
how this new formulation, which can be generalized to
many other rational conformal field theories, can be used
to compute correlation functions and to obtain new
expressions for the Virasoro and affine characters in
the theory.
}

\vfill


\newpage

\baselineskip=17pt

\newsection{Introduction}

Conformal Field Theory (CFT) in two dimensions has found
applications in a wide area of topics in
theoretical physics. In the context of this Summer School,
the first applications that come to mind are in
the area of String Theory, where CFT
techniques are an essential tool.
We would like to stress here, though, that there are a
variety of other topics, both within and outside
the realm of theoretical High Energy Physics, where
CFT has found important applications.

One of the surprising developments of the last few
years has been the realization (by many) that there are
often a variety of entirely different ways to set up
the description of a given CFT.
As an illustration, we mention that conformal
characters can often be written in different forms.
The fact that such different forms give rise to the
same $q$-series has led to non-trivial combinatorical
identities generalizing the so-called Rogers-Ramanujan
identities. A particularly interesting observation was
made in \cite{hhtbp}, where a new formulation of the
$SU(2)$, $k=1$, Wess-Zumino-Witten (WZW) CFT was proposed.
This new formulation does not rely on the affine Kac-Moody
and Virasoro symmetries, but instead uses invariance
under the Yangian $Y(sl_2)$ to give a systematic description
of the spectrum in terms of so-called multi-spinon states.

The spinon formulation of the $SU(2)_1$ WZW model has been
further worked out in \cite{bps,bls}, where it was also
observed that the new formulation can actually be used
to derive alternative expressions for both the affine and
the Virasoro characters. These papers thus established a
connection with some of the character identities that had
been proposed on different grounds and explained their origin.

It is the purpose of these lectures to explain in some
detail the new formulation of the $SU(2)_1$ WZW model
and to compare this with the conventional approach.
We have chosen to focus entirely on this one specific
(and well-studied) theory, hoping that this will help to
clarify the new issues that we address. Thus we shall only
briefly touch on one of the most interesting questions
in this field, which is: in what way can our results be
extended to more general rational conformal field theories
and perturbations thereof? New results in this direction
will be published elsewhere \cite{bls2,tbp}.

The organization of these lectures will be as
follows. In section~2, we review some basic features of
CFT in general and of the $SU(2)_1$ WZW model in
particular. In section~3 we derive the celebrated
Knizhnik-Zamolodchikov (KZ) equations and solve them
for the case of the 4-spinon correlation function.
In section~4 we focus on the spinon fields and
derive generalized commutation relations for their
Fourier modes. In section~5 we present the complete
spinon formulation $SU(2)_1$ WZW model by specifying
the rules for building a complete set of independent
multi-spinon states. In section~6 we use spinon
formulation to compute multi-spinon correlation functions
and to derive new expressions for the Virasoro and affine
characters in the theory. Appendix A gives a brief
summary of the definition of the Yangian $Y(sl_2)$

The central event of the last week of this Trieste
Summer School was the presentation to P.~van Nieuwenhuizen
of one of the 1994 Dirac Medals, honoring his contribution
to the discovery of supergravity in 1976. This year's other
recipients, S.~Ferrara and D.Z.~Freedman, had received
their medals on earlier occasions. We would like to take
this opportunity to congratulate the laureates with this
long-due token of appreciation for their monumental work.

\newpage

\newsection{The $SU(2)$, level $k=1$, WZW model}

The $SU(2)_1$ WZW model is a particularly simple CFT,
which can be used nicely to illustrate a
`quasi-particle approach' to CFT. In this case, the
quasi-particles are spinons, which, as we shall show,
can be viewed as free particles that obey semionic
statistics. Before we come to the spinon formulation,
we shall review the conventional formulation of the
theory, which uses the affine algebra $A_1^{(1)}$
as its starting point. We shall in particular
focus on the structure of the Hilbert space of
physical states.

Being a conformal field theory, the $SU(2)_1$ WZW
model possesses conformal invariance, with the corresponding
currents $T(z)$ satisfying the operator product expansion
(OPE)
\be
  T(z)T(w) = { \thalf \over (z-w)^4}
             + { 2 \, T(w) \over (z-w)^2}
             + { \del T(w) \over (z-w) } + \ldots
\ee
The modes $L_n$, defined through $T(z)=\sum_n L_n z^{-n-2}$,
satisfy the Virasoro algebra
\be
  [ L_m, L_n ] = {\textstyle 1 \over 12} (m^3-m) \delta_{m+n}
     + (m-n) L_{m+n}
\ee
with central charge $c=1$. The full chiral spectrum of the
theory can be viewed as a collection of irreducible
representations of the Virasoro algebra, each characterized
by a conformal dimension $h$. There is actually
an infinite number of such representations $L_j$,
with $h$ taking the values $h_j = j^2$, $j=0,\thalf,1 ,\ldots$.
In this notation, $j$ indicates the $SU(2)$ spin of
the highest weight state of the representation.

The number of states at a given $L_0$-level in each of
the Virasoro representations $L_j$ is summarized in the
characters
\be
\chi^{\rm Vir}_{j^2}(q) =  {\rm tr}_{L_j}(q^{L_0})
   = {q^{j^2}( 1 - q^{2j+1}) \over
     \prod_{m=1}^\infty (1-q^m)} \ .
\label{viraj}
\ee
The structure of these characters is easily
understood as follows. The product in the denominator
comes from the repeated application of the (bosonic)
Virasoro modes $L_{-m}$, $m=1,2,\ldots$, and
the subtraction in the numerator corresponds to
a null state of dimension $L_0=(j+1)^2$ in the
representation $L_j$. For example, for $j=0$
we have the null state $L_{-1}\vac$ with
$L_0=1$.

A more economical description is obtained by exploiting the
invariance of this field theory under the affine Lie
algebra $A_1^{(1)}$ with level $k=1$. The corresponding
currents $J^a(z)$ satisfy the OPE
\be
J^a(z)J^b(w) = {d^{ab} \over (z-w)^2}
               + {f^{ab}{}_c J^c(w) \over (z-w)}
               + \ldots
\ee
The adjoint index takes the values
$a=\pp,3,\mm$; the metric is $d^{\pp\mm}= 1$,
$d_{\mm\pp} = 1$, $d^{33} =2$, $d_{33}= \thalf$ and
the structure constants follow from $f^{\pp\mm}{}_{3}= 1$.
The modes $J_m^a$, defined through $J^a(z) =
\sum_n J_n^a z^{-n-1}$, satisfy the algebra
\be
  [ J^a_m , J^b_n ] = m\, d^{ab} \delta_{m+n}
             + f^{ab}{}_c J^c_{m+n} \ .
\label{a11}
\ee
The full chiral spectrum of the field theory consists
of two irreducible representations of the algebra
(\ref{a11}), the corresponding primary fields
being the identity and the field $\phi^{\alpha}(z)$,
of conformal dimension $h=\tquart$,
which we will call the {\it spinon field}.
The index $\alpha$ is an $SU(2)$ doublet index,
taking values $+,-$, and we have the OPE
\be
J^a(z)\phi^\alpha(w) = (t^a)^\alpha{}_\beta
                       {\phi^\beta(w) \over (z-w)}
                       + \ldots \ ,
\ee
with $(t^{\pp})^-{}_+ = (t^{\mm})^+{}_- = 1$,
$(t^3)^\pm{}_\pm =  \pm 1$.

The characters of the two representations
(labeled by $j=0,\thalf$) are given by
\bea
\chi_j^{A_1^{(1)}}(q) &=&
  {q^{-{1 \over 12}} \over \prod_{m=1}^{\infty} (1-q^m)^3}
  \sum_{n\in \ZZ} {\textstyle (6n+2j+1)} q^{(6n+2j+1)^2 \over 12}
\nonu
  &=&
  {q^{j^2} \over \prod_{m=1}^{\infty} (1-q^m)^3}
  \left[ {\textstyle (2j+1)} - {\textstyle (5-2j)}
  q^{2-2j} + \ldots \right] \, .
\label{affj}
\eea
where the dots indicate the subtraction of
further null states. These character formulas can
be understood as follows: the product in the
denominator accounts for the presence of
bosonic modes $J_m^a$ and the subtractions
correspond to null states in the affine modules.
The simplest null states (corresponding to the first
correction term in (\ref{affj})) are given by
\bea
  j=0 \, :  && J^a_{-1} J^b_{-1}
               |\, 0 \, \rangle |_{j=2\; {\rm component}}
\nonu
  j=\thalf \, :  && J^{a}_{-1} |\, \phi^\alpha\, \rangle
                   |_{j=\threehalves\; {\rm component}}
\label{nullstates}
\eea
where $|\, \phi^\alpha\, \rangle$ denotes the highest
weight state of the $j=\thalf$ representation, and the
subscripts on the states denote projections onto
irreducible representations of $SU(2)$.

An alternative form of the affine characters is as follows
\be
\chi_j^{A_1^{(1)}}(q) =
 {\sum_{n\in \ZZ} q^{(n+j)^2} \over \prod _{m=1}^\infty
  (1-q^m)} \ .
\ee
This form is natural from the point of view of a
representation of the $SU(2)_1$ theory in terms of a single
bosonic field (compare with section~5.2, where we shall use
this bosonic field in the construction of Yangian Highest
Weight Vectors).

\newpage

\newsection{Analyzing the KZ equations}

Having discussed the spectrum of physical states
we now turn to correlation functions of the
fundamental spinon field $\phi^\alpha(z)$.
We first observe that for $k=1$ the existence
of the second null state in (\ref{nullstates})
implies that
\be
   (J^a\phi^\alpha)(z) = 2 (t^a)^{\alpha}{}_{\beta}
                         \del \phi^\beta(z) \ .
\label{nullfield}
\ee
Here we use ordinary brackets to denote the
normal ordered field product defined by
\be
     (AB)(z) = \oint_{{\cal C}_z} {dx \over 2\pi i}
                  {A(x) B(z) \over (x-z)} \ ,
\ee
where the contour ${\cal C}_z$ encloses the point
$z$ counterclockwise.
{}From (\ref{nullfield}) the following identity
can be derived
\be
2 \, (t^a)^{(i)} \, \del_{z_i} \,
   \langle \phi(z_1) \ldots \phi(z_N) \rangle
= \sum_{j\neq i}
   {(t^a)^{(j)} \over z_i-z_j } \,
   \langle \phi(z_1) \ldots \phi(z_N) \rangle \ ,
   \label{KZ1}
\ee
where we suppressed spinor indices.
For general level $k$ we have the following equation,
which is known as the Knizhnik-Zamolodchikov (KZ)
equation \cite{KZ}
\be
0 =
\left( (k+2) \, \del_{z_i} - \sum_{j\neq i}
       { (t_a)^{(i)} (t^a)^{(j)} \over z_i-z_j } \right)
   \langle \phi(z_1) \ldots \phi(z_N) \rangle \ .
\label{KZ}
\ee
For $k=1$ the equation (\ref{KZ}) follows from (\ref{KZ1})
by taking a contraction.

We shall now derive an explicit expression for the
4-point function
\be
  G_{\alpha_1 \alpha_2}^{ \alpha_3 \alpha_4}(z_1,z_2,z_3,z_4)
  = \langle \phi_{\alpha_1}(z_1) \phi_{\alpha_2}(z_2)
       \phi^{\alpha_3}(z_3) \phi^{\alpha_4}(z_4) \rangle \ ,
\ee
by using conformal invariance and the equation (\ref{KZ})
for $k=1$. (We raise and lower spinor indices according
to $X^\alpha = \epsilon^{\alpha\beta} X_\beta$,
$X_\alpha = \epsilon_{\alpha\beta} X^\beta$,
where $\epsilon^{+-}=-\epsilon^{-+}=1$, $\epsilon_{-+}
=-\epsilon_{+-}=1$.)
Conformal invariance alone tells us that the 4-point
function can be written as
follows
\be
  G_{\alpha_1 \alpha_2}^{\alpha_3 \alpha_4}(z_1,z_2,z_3,z_4)
  = (z_{14} z_{23})^{-\half}
  \, G_{\alpha_1 \alpha_2}^{\alpha_3 \alpha_4}(x) \ ,
\ee
where we wrote
\be
  x = { z_{12} z_{34} \over z_{14} z_{32} } \ , \qquad
  z_{ij} = z_i-z_j \ .
\ee
By exploiting $SU(2)$ invariance we obtain the following
general form of the solution
\be
  G_{\alpha_1 \alpha_2}^{ \alpha_3 \alpha_4}(x) =
    (J_1)_{\alpha_1 \alpha_2}^{ \alpha_3 \alpha_4} \, G_1(x)
    + (J_2)_{\alpha_1 \alpha_2}^{\alpha_3 \alpha_4} \, G_2(x) \ ,
\ee
where
\be
(J_1)_{\alpha_1 \alpha_2}^{\alpha_3 \alpha_4}
= \epsilon_{\alpha_1 \alpha_2} \epsilon^{\alpha_3 \alpha_4}
\ , \qquad \qquad
(J_2)_{\alpha_1 \alpha_2}^{\alpha_3 \alpha_4}
= \delta_{\alpha_1}^{ \alpha_3} \delta_{\alpha_2}^{\alpha_4} \ .
\ee
Using the identities (suppressing indices)
\bea
(t_a)^{(1)} (t^a)^{(2)} J_1 = - \threehalves J_1 \ ,
&& \qquad
(t_a)^{(1)} (t^a)^{(2)} J_2 = - J_1 + \thalf J_2 \ ,
\nonu
(t_a)^{(1)} (t^a)^{(3)} J_1 = \thalf J_1 - J_2 \ ,
&& \qquad
(t_a)^{(1)} (t^a)^{(3)} J_2 = - \threehalves J_2 \ ,
\nonu
(t_a)^{(1)} (t^a)^{(4)} J_1 = - \thalf J_1 + J_2 \ ,
&& \qquad
(t_a)^{(1)} (t^a)^{(4)} J_2 = J_1 - \thalf J_2 \ ,
\eea
one can rewrite the KZ equation (\ref{KZ}) into the
following differential equation
\be
  -6 \del_x  \left( \begin{array}{c} G_1(x) \\ \stru G_2(x)
                     \end{array} \right)
  = \left[ {1 \over x} \left( \begin{array}{cc} 3 & 2 \\ 0 & -1
                       \end{array} \right)
    + {1 \over x-1} \left( \begin{array}{cc} -1 & 0 \\ 2 & 3
                    \end{array} \right) \right]
    \left( \begin{array}{c} G_1(x) \\ \stru G_2(x)
                     \end{array} \right) \ .
\ee
The solution to these equations takes the simple form
\be
    G_1(x) = (1-x)^{\half} x^{-\half} \ , \qquad
    G_2(x) = (1-x)^{-\half} x^{\half} \ .
\label{KZsol}
\ee
Obviously, the final form of the correlation function
could have been obtained by using that the $SU(2)_1$ WZW
model is a $c=1$
CFT that is equivalent to a theory of a single real
scalar field compactified on a circle of the appropriate
radius. However, we presented this derivation
to illustrate how the presence of affine symmetry can
be exploited to solve for interesting quantities such
as correlation functions. In section~6 we shall present
yet another derivation of this same result, that time
using the systematics of the spinon formulation.

Before closing this section, we would like to digress and
show how the result (\ref{KZsol}) for the spinon 4-point
function can be used to gain some insight in the structure
of the Hilbert space, in particular about 2-spinon states.

To this end we map the complex $z$-plane
into the cylinder, $z_i= \exp{(-{2\pi \over l}w_i)}$,
and write $w_i=\tau_i+ix_i$. Under this conformal
transformation, the 4-point function
picks up a factor $\prod_i (dz_i/dw_i)^{1 \over 4}$.
We impose time ordering on the $\tau_i$, choosing
$\tau_{i+1}-\tau_i=0^+$, and pass to a
description using modes $\phi_n^\alpha$
defined through
\be
    \phi^\alpha(x,0) =
    \sum_n \phi_n^\alpha \, e^{i{2\pi \over l}nx} \ .
\ee
With that, the 4-point function can be
written as
\be
G(x_1,x_2,x_3,x_4) = \sum_{n_1,n_2,n_3,n_4}
  e^{ i {2\pi \over l} ( n_1 x_1 + n_2 x_2 + n_3 x_3 + n_4 x_4) }
  \langle \, 0 \, | \phi_{n_1} \phi_{n_2}
  \phi_{n_3} \phi_{n_4}\, | \, 0 \, \rangle \ .
\ee
By decomposing the explicit result
(\ref{KZsol}) in a series, we can thus extract exact
information on the inner products of pairs
of 2-spinon states. Explicitly, we have
\bea &&
G_1(x_1,x_2,x_3,x_4)  =
  e^{{2\pi \over l}{1 \over 4} ( x_1-x_2+x_3-x_4)}
  \, (1 + \thalf e^{{2\pi \over l} x_{23}} + \ldots) \ ,
\nonu &&
G_2(x_1,x_2,x_3,x_4)  = -
  e^{{2\pi \over l}{1 \over 4} ( x_1+ 3 x_2 - 3 x_3 - x_4)}
  \, (1 + \thalf e^{{2\pi \over l} x_{23}} + \ldots) \ .
\eea
If we now, for instance, project on the triplet channel
for the indices $\alpha_3,\alpha_4$, implying
that we consider $G_2$ only, we find that
the following 2-spinon states are nonvanishing
\be
  (t^a)_{\alpha\beta}  \, \phi^\alpha_{-{3 \over 4}}
                       \phi^\beta_{-{1 \over 4}}
                       \, | \, 0 \, \rangle \ , \qquad
  (t^a)_{\alpha\beta}  \, \phi^\alpha_{-{7 \over 4}}
                       \phi^\beta_{-{1 \over 4}}
                       \, | \, 0 \, \rangle \ , \qquad {\rm etc.}
\ee
Similarly, in the singlet channel, where $(2G_1+G_2)$
is the relevant combination, we find the following
2-spinon states
\be
  \epsilon_{\alpha\beta} \, \phi^\alpha_{{1 \over 4}}
                       \phi^\beta_{-{1 \over 4}}
                       \, | \, 0 \, \rangle \ , \qquad
  \epsilon_{\alpha\beta}  \, \phi^\alpha_{-{7 \over 4}}
                       \phi^\beta_{-{1 \over 4}}
                       \, | \, 0 \, \rangle \ , \qquad {\rm etc.}
\ee
We would like to give two comments on these results.
First, one notes that the mode-index of the
second $\phi$ factor is shifted from $(-\tquart \,{\rm mod}\, \ZZ)$
to $(-{3 \over 4}\, {\rm mod}\, \ZZ)$. Second, one observes that the
state with mode-indices $(-{3 \over 4},-\tquart)$ vanishes
in the singlet channel. The vanishing of specific multi-spinon
states can be understood from a `generalized Pauli principle' for
semionic operators, which is to be discussed in sections 4 and 5.

\newpage

\newsection{Spinon formulation}

\vspace{-8mm}

\newsubsection{Generalized commutation relations}

In this section we start our discussion of what we
call the `spinon formulation' of the $SU(2)_1$ WZW model.
Since we shall be considering the repeated application of
modes of the spinon fields $\phi^\alpha(z)$, we start by
deriving generalized commutation relations that are satisfied
by these modes.

Our starting point are the following spinon-spinon OPE's
\bea
\phi^\alpha(z)\phi^\beta(w) &=&
    (-1)^q (z-w)^{-\half} \epsilon^{\alpha\beta}
    \left( 1 + \thalf (z-w)^2 T(w) + \ldots \right)+
\nonu
&& - (-1)^q (z-w)^{\half} (t_a)^{\alpha\beta}
    \left( J^a(w) + \thalf (z-w) \del J^a(w) + \ldots \right) \ ,
\eea
where $(t_a)^{\alpha\beta} = d_{ab}
\epsilon^{\beta\gamma} (t^b)^\alpha{}_\gamma$.
In these formulas, $q$ depends on the sector that the
OPE's are acting on : $q=0$ on states that are created
by an even number of spinons (\ie, the states in the
vacuum module of $A_1^{(1)}$)
and $q=1$ if the number of spinons is odd (the $j=\thalf$
module of $A_1^{(1)}$).
The appearance of explicit factors
$(-1)^q$ is due to our convention to write upper indices
on all spinon fields. The natural convention would be
to write lower indices whenever a spinon field acts on
a $q=1$ state; raising these indices leads to a
relative minus sign between the OPE's in the two sectors
$q=0,1$.

The occurence of a factor $(z-w)^{1/2}$ in the
spinon-spinon OPE's clearly shows that the braiding
properties of the spinons are those of semions or
`half-fermions.'
The mode expansions of the spinons  are
\bea
&& \phi^{\alpha}(z)\chi_q(0) = \sum_m z^{m+\qq}
\phi^{\alpha}_{-m-\qq-{1 \over 4}}
  \chi_q(0) \ ,
\nonu
&& \phi^{\alpha}_{-m-\qq+{3 \over 4}} \chi_q(0) =
\oint {dz \over 2 \pi i} z^{-m-\qq}
\phi^{\alpha}(z)\chi_q(0) \ ,
\eea
where $\chi_q(0)$ is an arbitrary state in the sector indicated
by the value of $q$. On states with $q=0$ we can apply modes
$\phi^\pm_{-1/4-n}$ with $n$ integer, and on states with
$q=1$ we can apply $\phi^\pm_{-3/4-n}$.

By following a standard procedure (see \eg\ \cite{faza})
we can derive the following relations for the modes of the
spinon fields
\be
\sum_{l\geq0} C_l^{(-{1 \over 2})}
   \left( \phi^\alpha_{-m-{q+1 \over 2}-l+{3 \over 4}}
          \phi^\beta_{-n-\qq+l+{3 \over 4}}
   - \left(
\begin{array}{c}
 \alpha  \leftrightarrow \beta
\\
m \leftrightarrow n
\end{array} \right) \right)
   = (-1)^q \, \epsilon^{\alpha\beta} \, \delta_{m+n+q-1} \ ,
\label{gc1}
\ee
\bea
&& \sum_{l\geq0} C_l^{(-{3 \over 2})}
   \left( \phi^{\alpha}_{-m-{q+1 \over 2}-l-{1 \over 4}}
          \phi^{\beta}_{-n-\qq+l+{3 \over 4}}
   + \left(
\begin{array}{c}
 \alpha  \leftrightarrow  \beta
\\
 m \leftrightarrow n
\end{array} \right) \right)
\nonu
&& \qquad \qquad = (-1)^q \left( - \epsilon^{\alpha\beta} (m+\qq)
   \delta_{m+n+q} - (t_a)^{\alpha\beta} J^a_{-m-n-q} \right) \ ,
\label{gc2}
\\[2mm]
&& \sum_{l\geq0} C_l^{(-{5 \over 2})}
   \left( \phi^\alpha_{-m-{q+1 \over 2}-l-{5 \over 4}}
          \phi^\beta_{-n-\qq+l+{3 \over 4}}
   - \left(
 \begin{array}{c}
\alpha  \leftrightarrow \beta
\\
m \leftrightarrow n
 \end{array} \right) \right)
\nonu
&& \qquad\qquad = (-1)^q \left( \thalf \,
     \epsilon^{\alpha\beta} \,
     (m+\qq)(m+1+\qq) \delta_{m+n+q+1} \right.
\nonu
&& \qquad\qquad\qquad\qquad \left.
 - \thalf (t_a)^{\alpha\beta} (n-m) J^a_{-m-n-q-1}
 + \thalf \, \epsilon^{\alpha\beta} L_{-m-n-q-1} \right) \ .
\label{gc3}
\eea
In these relations, the coefficients $C_l^{(\alpha)}$ are defined
by the expansion
\be
(1-x)^\alpha = \sum_{l\geq 0} C_l^{(\alpha)} x^l \ .
\ee
Algebras of the type (\ref{gc1}) are known as
`generalized vertex algebras' \cite{dong}; other examples
include the so-called parafermion algebras \cite{faza} or
$Z$-algebras \cite{lewi}.

The relations (\ref{gc1}) can be interpreted
as {\it generalized canonical commutation relations} of the
fundamental spinon fields. The other relations can be used to
express the current modes $J^a_n$ and the Virasoro generators
$L_n$ as bilinears in spinon modes. We would like to stress
that, up to the complication of the infinite series in the
mode index $l$, these relations are very similar to the
anticommutation relations for free fermions and to the
formulas that express affine and Virasoro currents as fermion
bilinears.

With the generalized commutation relations (\ref{gc1}) in
place, we are now ready to consider the Fock space generated
by the spinon fields acting on the vacuum. By this we mean that
we construct all possible multi-spinon states, taking into
account equivalences that follow from (\ref{gc1}). A
surprising result, which we shall further discuss in the
next section, is the following
\begin{quote}
{\em The (chiral part of the) Hilbert space of the
     $SU(2)_1$ WZW model
     is identical to the spinon Fock space
     constructed using the generalized commutation
     relations (\ref{gc1}).}
\end{quote}
To illustrate what this result really means, let
us look back at the formulation of the same theory
using affine Kac Moody algebras (see section~2). In
that formulation, one constructs affine modules
by applying affine current modes $J_m^a$ to a highest
weight state (giving `current Fock spaces' as opposed to
`spinon Fock spaces'). An important point was that in
those affine modules there were null states, \ie, states
whose vanishing does not directly follow from the commutation
relations of the $J$'s. (Examples of such null states are
the states (\ref{nullstates}). In the spinon
formulation of the theory the only relations among the
multi-spinon states are those implied by the generalized
commutation relations (\ref{gc1}), which express
the semionic nature of the fields. There are no
null states, and the theory can be viewed as a
theory of {\it free}\ spinon fields.

As a comment, we mention that there are many more
conformal field theories for which the
Hilbert space can be generated by repeatedly
acting with modes of one or a few low-dimension primary
fields. However, in general one should expect that the
Fock space generated from such generalized spinons
is larger that the actual Hilbert space of the theory.
The systematics of reducing such Fock spaces to the
true Hilbert space have not been worked out. What is
so special about the $SU(2)_1$ WZW model and about a
number of other examples \cite{bls2,tbp}, is that the Hilbert
space is actually a free Fock space of fractional
statistics objects.

\newsubsection{Yangian symmetry}

When setting up a description of the $SU(2)_1$
WZW field theory in terms of multi-spinon states,
one soon discovers that the affine and Virasoro
symmetries are not very convenient tools, the
problem being that their action on multi-spinon
states is not easily tractable.
Interestingly, it has been found that this same
field theory admits another highly non-trivial
symmetry structure, the so-called Yangian
$Y(sl_2)$, which is very natural from the point
of the spinon formulation.

In Appendix A we recall the definition of the
Yangian $Y(sl_2)$, which is an example of a non-trivial
quantum group. Let us now show that this algebra
can be represented on the Hilbert space of the
$SU(2)_1$ WZW model. Following \cite{hhtbp}, we make
the following definitions
\be
Q_0^a = J_0^a \, , \quad Q_1^a = \thalf f^a{}_{bc} \sum_{m>0}
  J_{-m}^b J_m^c \ .
\label{yn}
\ee
It may be checked that,
when acting on integrable highest weight representations
of $A^{(1)}_1$ at level $k=1$, these generators satisfy the
defining relations (\ref{terrific}) of $Y(sl_2)$.

It is easily seen that the Virasoro generator $L_0$ commutes
with the Yangian generators (\ref{yn}). It is actually
expected \cite{hhtbp,ht,bps} that there exists an infinite number
of mutually commuting operators $H_n$, $n=1,2,\ldots$, that all
commute with the Yangian, the first few examples being $H_1=L_0$
and
\be
H_2 = d_{ab} \, \sum_{m>0} \, m \, J^a_{-m} J^b_m \ .
\label{H2}
\ee

The existence of Yangian symmetry and of the operators
$H_n$ suggests a description of the Hilbert space
in terms of irreducible multiplets of the Yangian,
which are each characterized by specific eigenvalues
of the $H_n$. As we shall see, such multiplets are
naturally described in terms of multi-spinon states.

As an aside we remark that, if the group $SU(2)$
is replaced by $SU(N)$, the expressions (\ref{yn})
and (\ref{H2}) for $Q_1^a$ and $H_2$ have to be modified
by additional terms that involve the 3-index $d$-symbol
of $SU(N)$, $N\geq 3$~\cite{kjs}.

\newpage

\newsection{Structure of the Hilbert space}

\vspace{-8mm}

\newsubsection{Example: 2-spinon states}

To illustrate the connection between the spinon
formulation and Yangian symmetry, we shall
first analyze in some detail the structure of the
2-spinon states in the spectrum.

We introduce the following notations for general
2-spinon states ($t$, $s$ refer to $SU(2)$ triplet
and singlet channels, respectively)
\begin{equation}
  \Phi_{n_2,n_1}^{t,a} =
  (t^a)_{\alpha\beta}  \ \phi^\alpha_{-3/4-n_2}
                         \phi^\beta_{-1/4-n_1} \vac,
\quad
  \Phi_{n_2,n_1}^s =
  \epsilon_{\alpha\beta} \ \phi^\alpha_{-3/4-n_2}
                            \phi^\beta_{-1/4-n_1} \vac \ .
\end{equation}
The energy eigenvalue of these states is simply
\be
 L_0 = 1 + n_1 + n_2 \ .
\label{Lze}
\ee
We can now use eq.~(\ref{nullfield}) to show that
(\ref{yn}) leads to
\bea
&& Q_1^a \Phi_{n_2,n_1}^{t,b} =
  -(n_2+n_1+\thalf) f^{ab}{}_c \Phi_{n_2,n_1}^{t,c}
  + (n_2-n_1+1) d^{ab} \Phi_{n_2,n_1}^s
    + d^{ab} \sum_{l>0}  \Phi_{n_2+l,n_1-l}^s
\nonu
&& Q_1^a \Phi_{n_2,n_1}^s =
  2 (n_2-n_1) \Phi_{n_2,n_1}^{t,a}
  - 2 \sum_{l>0} \Phi_{n_2+l,n_1-l}^{t,a} \ .
\eea
The first thing to notice from these formulas is the fact
that the Yangian generators $Q_1^a$ map 2-spinon
states into 2-spinon states. This illustrates
our claim that the Yangian is entirely natural
from the point of view of the spinon formulation.
Notice also that the action of $Q_1^a$ is not diagonal
in the indices $(n_2,n_1)$ but rather lower-triangular
in the sense that $(n_2,n_1)$ gets mapped into
$(n_2+l,n_1-l)$ with $l\geq 0$ and $n_1 - l \geq 0$.

{}From the action of $Q_1^a$ it is easily seen that
the space of all two-spinon states with $n_1+n_2=n$ fixed
can be decomposed into multiplets of the
Yangian. Each multiplet contains a Yangian Highest
Weight Vector (YHWV), \ie, a state that is highest weight
with respect to $SU(2)$, is an eigenvector for $Q_1^3$ and
is annihilated by $Q_1^{\pp}$. These YHWV's are of the form
\be
\Phi_{n_2,n_1}^{t,\pp} + \sum_{l>0} a_{n_2,n_1}^{(l)}
    \Phi_{n_2+l,n_1-l}^{t,\pp} \ ,
\label{2spYHWV}
\ee
where the $a_{n_2,n_1}^{(l)}$ are real coefficients.
The 2-spinon Yangian multiplets each contain a
triplet and a singlet of $SU(2)$, \ie, a total of four
states, except if $n_1=n_2$, when
the relation (\ref{gc1}) can be used to show
that the singlet is absent.

We remark that  $Q_1^a$  acts by comultiplication
(\ref{copr}) on the 2-spinon YHWV, given
its action on the 1-spinon states (which are YHWV).

Turning to the operator $H_2$, using
the relation
\be
   (\del J^a\phi^\alpha)(z) = {4 \over 3}
          (t^a)^\alpha{}_\beta \del^2 \phi^\beta(z) \ ,
\label{nfs}
\ee
we find that the action
on 2-spinon states is as follows
\bea
H_2 \Phi_{n_2,n_1}^{t,a} &=&
 2\left( (n_2+1)(n_2+\half) + (n_1+\half)n_1\right)
    \Phi_{n_2,n_1}^{t,a}
 + \sum_{l>0} \, l \, \Phi_{n_2+l,n_1-l}^{t,a} \ ,
\nonu
H_2 \Phi_{n_2,n_1}^s &=&
 2\left( (n_2+1)(n_2+\half) + (n_1+\half)n_1\right)
    \Phi_{n_2,n_1}^s
 - 3 \, \sum_{l>0} \, l \, \Phi_{n_2+l,n_1-l}^s \ .
\nonu
\eea
This action is lower triangular, and the
eigenvalues of the operator $H_2$ are immediately
seen to equal
$2\left( (n_2+1)(n_2+\half) + (n_1+\half)n_1\right)$.
As $H_2$ commutes with the Yangian, the $H_2$ eigenstates
group into Yangian multiplets, and the YHWV's are given
by the $H_2$ eigenstates with triplet index $\pp$.

To be completely explicit, we present the example
where $n_1+n_2=4$, which are the 2-spinon states with
$L_0=5$ ( from (\ref{Lze})).
In the following formula we list the labels $(n_2,n_1)$
of the Yangian representation, the $H_2$-eigenvalues, and
the states
\bea
&& (4,0) \qquad H_2=45 \qquad
   \Phi_{4,0}^{t,a}\ , \quad \Phi_{4,0}^s
\nonu
&& (3,1) \qquad H_2=31 \qquad
   \Phi_{3,1}^{t,a} - \ts{\frac{1}{14}} \Phi_{4,0}^{t,a}\ , \quad
   \Phi_{3,1}^s + \ts{\frac{3}{14}} \Phi_{4,0}^s
\nonu
&& (2,2) \qquad H_2=25 \qquad
  \Phi_{2,2}^{t,a} - \ts{\frac{1}{6}} \Phi_{3,1}^{t,a}
       - \ts{\frac{11}{120}} \Phi_{4,0}^{t,a} \ .
\label{mults}
\eea
When acting on the $(2,2)$ YHWV,
$Q_1^{\mm}$ produces a multiple of the $SU(2)$ descendant
plus a state proportional to
$\Phi_{2,2}^s+ \half \Phi_{3,1}^s + \frac{3}{8} \Phi_{4,0}^s$,
which vanishes as a consequence of the commutation relation
(\ref{gc1}).

\newsubsection{$N$-spinon states}

The analysis of the previous subsection can
in principle be generalized to the case of general
$N$-spinon states. We shall first focus on the
following set of states, which we call
{\it fully polarized $N$-spinon states}
\bea
&&    \phi^+_{-{(2N-1) \over 4}-n_N}  \ldots
      \phi^+_{-{5 \over 4}     -n_3}
      \phi^+_{-{3 \over 4}     -n_2}
      \phi^+_{-{1 \over 4}     -n_1} |0 \rangle \, ,
\nonumber \\[4mm]
&& \qquad {\rm with} \;\; n_N \geq n_{N-1} \geq \ldots
    \geq n_2 \geq n_1 \geq 0 \ .
\label{pol}
\eea
The eigenvalue of the
Virasoro zero mode $L_0$ on these states is
\be
   L_0 = {N^2 \over 4} + \sum_{i=1}^N n_i \ .
\label{Lzero}
\ee
One easily derives the following explicit expression
for the action of $H_2$ on a fully polarized $N$-spinon
state. Denoting by $|\chi^{(N-1)}\rangle$ a fully
polarized $(N-1)$-spinon state, we have
from (\ref{H2}), (\ref{nfs})
\bea
\lefteqn{ [ H_2, \phi^+_{-{2N-1 \over 4}-n_N} ]
    |\chi^{(N-1)}\rangle =}
\nonu && \!\!\!\!\!
2 \, (n_N+ {\ts {N-1 \over 2}}  )
     (n_N+ {\ts {N \over 2}}  ) ) \,
    \phi^+_{-{2N-1 \over 4}-n_N}
    |\chi^{(N-1)}\rangle
 + \sum_{l>0} \, l \, \phi^+_{-{2N-1 \over 4}-n_N-l} J^3_l
    |\chi^{(N-1)}\rangle  \ .
\nonu
\eea
Since this action is again `lower triangular,'
the eigenvalue of $H_2$ on the eigenstate
labeled by mode-indices $\{n_1,n_2,\ldots,n_N\}$
is found to be
\be
  H_2 = \sum_{i=1}^N \, 2 \, (n_i+\thalf(i-1))(n_i+\thalf i) \ .
\ee

The fully polarized $N$-spinon states that are eigenstates
of $H_2$ are YHWV's. By
acting with the Yangian generators $Q_0^a$ and $Q_1^a$ we
may construct Yangian multiplets. Obviously, all
the states in one such multiplet share common
eigenvalues for the operators $H_n$, $n=1,2,\ldots$.

The precise structure of the Yangian multiplets
has first been explored in the context of the
so-called Haldane-Shastry spin chain with
inverse square exchange \cite{hhtbp}. The $SU(2)_1$
conformal field theory can be viewed as
a continuum limit of the Haldane-Shastry
chain, and many results, in particular statements
about Yangian symmetry and about the structure of the
spectrum, carry over to the field theory. Backing
all these results is the representation theory
of Yangians, which has been worked out in detail
in \cite{cp}.

In the language that we developed, the Yangian multiplets
can be characterized as follows.
(i): Each Yangian multiplet is characterized by a set of
non-decreasing integers
${ \{ n_i \}}_{i=1,...,N}$
as in
 (\ref{pol}).
 (ii): The eigenvalue  of $L_0$ ( commuting
with the Yangian ) on the
states in the Yangian multiplet specified by
${ \{ n_i \}}_{i=1,...,N}$  is given by (\ref{Lzero}).
(iii): When acting on a YHWV (which  will be a linear
combination of
fully polarized states of the form (\ref{pol})),
the Yangian generators (\ref{yn})
create states of a similar form, which however
have some of the
$+$ indices replaced by $-$, and which have different
coefficients in the linear combination. If we were to
act only with $Q_0^a$ we would find a total of $N+1$ such states;
the maximal possible number when acting with the full Yangian
is $2^N$. This maximal number is only realized if
the mode-indices $n_i$ are all different. If some of the $n_i$'s
are equal, the corresponding product
of doublets is projected on the symmetric combination. For
example, a 2-spinon Yangian multiplet will have $3+1=4$ states
if $n_2\neq n_1$, but only 3 states if $n_2=n_1$.
The vanishing of some of the singlet channel
states is encoded in the generalized commutation
relations (\ref{gc1}).

The union of all Yangian multiplets, whose structure we
just described, precisely forms a basis of the Hilbert
space of the $SU(2)_1$ WZW model. The `Yangian rules' for
the construction of multi-spinon states can thus be
interpreted as a generalized Pauli principle that
governs the filling of possible momentum states of the
fundamental spinons. As we already mentioned, the full set
of rules reflects the semionic nature of the spinons,
which can otherwise be viewed as free objects.
It is instructive to compare the Pauli priciple
for spinons with that for spinful fermions, where,
as is well-known, double occupancy of a given momentum
state leads to anti-symmetrization of internal
indices.

We would now like to give
explicit expressions for the fully polarized
$N$-spinon states that are eigenstates of $H_2$,
\ie, the YHWV's. These results, which were first given
in \cite{bps}, have been inspired by the machinery that has
recently been developed for the analysis of a variety
of exactly solvable quantum mechanical systems with
inverse square exchange. The explicit formulas involve
so-called Jack polynomials, whose properties have
been studied in the mathematical literature (see,
\eg, \cite{stan}).
We will here follow the conventions and normalizations
of \cite{bps}.

The construction in \cite{bps} uses an auxiliary
free field $\varphi(z)=q-ip\ln z + i \sum_{n\neq 0}
\alpha_n {z^{-n} \over n}$, in terms of which the
fundamental spinons can be written as
\be
\phi^{\pm}(z) = \, : e^{\pm i {1 \over \sqrt{2}} \varphi(z)} : \ .
\ee
The $N$-spinon YHWV with labels
$\{n\}=\{n_N,n_{N-1},\ldots,1\}$ can be written as
\be
  | \{ n \} \rangle = (-1)^{|n|} \, P_{ \{n'\} }^ {(-2)}
    (p_n) \,  e^{i{N\over\sqrt{2}}q} \, | 0 \rangle \ .
\label{yhw}
\ee
In this formula, the set $\{ n' \}$ is dual to $\{n\}$
in the sense of partitions or, equivalently, Young
tableaux and we wrote $|n|=\sum n_i$. The function
$P_{\{n'\}}^{(-2)}$ is a
Jack polynomial, whose arguments $p_n$ are set
equal to the oscillators of the auxiliary boson
field $\varphi(z)$,
\be
p_n= - \thalf \sqrt{2} \, \alpha_{-n} \ .
\ee

Let us do an example and show that the explicit 2-spinon
YHWV's that we discussed before are indeed reproduced by this
general formula. Choosing $n_2=2,n_1=1$, we have
\bea
  | \{2,1\}\rangle &=& - P^{(-2)}_{\{2,1\}}(p_n) \,
     e^{i\sqrt{2}q} \, | 0 \rangle
\nonu
  &=& - {\textstyle {2 \over 5}} (p_1^3-\thalf p_2 p_1
           - \thalf p_3) \, e^{i\sqrt{2}q} \, | 0 \rangle \ .
\label{expr1}
\eea
In the language that we used before we would have
written this YHWV as
\be
      \Phi^{t,\pp}_{2,1} - {\textstyle {1 \over 10}}
             \Phi^{t,\pp}_{3,0}
\label{expr2}
\ee
(compare with (\ref{mults})). Explicit algebra
gives
\bea
\Phi^{t,\pp}_{2,1} &=&
  - \left(  P^{(-2)}_{\{1,1\}} \, P^{(-2)}_{\{1\}}
    - \thalf P^{(-2)}_{\{1,1,1\}} \right)(p_n) \,
    e^{i\sqrt{2}q} \, | 0 \rangle
\nonu
\Phi^{t,\pp}_{3,0} &=&
   - P^{(-2)}_{\{1,1,1\}}(p_n) \, e^{i\sqrt{2}q} \, | 0 \rangle
\eea
and the two expressions (\ref{expr1}), (\ref{expr2})
for the YHWV are seen to agree by using the recursion
relation
\be
P^{(-2)}_{\{2,1\}} = P^{(-2)}_{\{1,1\}} P^{(-2)}_{\{1\}}
 - {\textstyle {3 \over 5}} P^{(-2)}_{\{1,1,1\}} \ .
\ee

The YHWV's satisfy the following orthogonality
condition
\be
  \langle \{ m \}   | \{ n \} \rangle =
              \delta_{ \{m\}, \{n\} } \, j_{\{n'\}} \ .
\ee
Explicit results for the metric $j_{\{n\}}$ may
be found for example in \cite{stan}.
In the next section, we will use these general
results for the explicit computation of
$2N$-point correlation functions.

Before closing this section, we present a second way
to write a  multi-spinon basis for the $SU(2)_1$ WZW
model. One considers the states
\bea
&&    \phi^-_{-{2(N^+ + N^-)-1 \over 4} -n^-_{N^-}}
      \ldots
      \phi^-_{-{2(N^+ +1)-1 \over 4}    -n^-_1}
      \phi^+_{-{2N^+ -1 \over 4}        -n^+_{N^+}}
      \ldots
      \phi^+_{-{1 \over 4}-n^+_1} |0\rangle \, ,
\nonumber\\[6mm]
&& \qquad {\rm with} \;\;
    n^+_{N^+} \geq \ldots \geq n^+_2 \geq n^+_1 \geq 0
    \ , \quad
    n^-_{N^-} \geq \ldots \geq n^-_2 \geq n^-_1 \geq 0 \ .
\label{plusmin}
\eea
Using an induction argument, it can easily be shown that
the generalized
commutation relations (\ref{gc1}) allow one
to write every mixed index multi-spinon state as a sum of
states of the form (\ref{plusmin}). The eigenvalue of $L_0$
is now given as
\be
   L_0 = { (N^+ + N^-)^2 \over 4}
         + \sum_{i=1}^{N^+} n^+_i
         + \sum_{i=1}^{N^-} n^-_i \ .
\ee

\newpage

\newsection{Applications}

\vspace{-8mm}

\newsubsection{Ward Identities and correlators}

In section~3 we discussed the KZ equations for
correlation functions in the $SU(2)_1$ CFT and
explicitly solved for the 4-spinon correlation
functions. In the spinon
formulation of the theory, one would like to
derive similar results, this time not by using
the KZ equations (which have their origin in the affine
symmetry of the theory) but rather by using
Ward identities that are directly related to the
Yangian symmetry and to the existence of the
higher conserved quantities $H_n$. Without
derivation (see however \cite{bps}), we present
the following Ward
identities, which are associated with $Q_1^a$
and $H_2$, respectively
\bea
&& 0= \left( -2 \, \sum_i (z_i \del_i) (t^a)^{(i)}
       + \thalf f^a{}_{bc} \sum_{i\neq j}
       \theta_{ij} (t^b)^{(i)} (t^c)^{(j)} \right)
   \langle \phi(z_1) \ldots \phi(z_N) \rangle \ ,
\nonu
&& 0= \left( 2 \, \sum_i ((z_i \del_i)^2
       + \thalf(z_i\del_i))
       - \sum_{i\neq j}
       \theta_{ij} \theta_{ji} (t_a)^{(i)}
       (t^a)^{(j)} \right)
   \langle \phi(z_1) \ldots \phi(z_N) \rangle \ .
\eea
It is easily seen that these two equations, together
with the Ward identities coming from translational
and scale invariance (\ie, $L_{-1}$ and $L_0$)
and from $Q_0^a$, are
sufficient to completely determine the 4-point
correlation functions. One expects that more general
correlators can be determined by invoking in addition
Ward identities coming from the higher conserved
quantities $H_n$, $n\geq 3$.
Rather than pursuing this road, we will now present a
direct computation of a general $2N$-spinon
correlation function.

Before we come to the correlation functions, let us
write an expression for the action of a product of spinon
fields $\phi^+(z)$ on the vacuum. We have \cite{bps}
\be
\phi^+(z_1) \ldots \phi^+(z_N) |0\rangle
=
\prod_{i<j} (z_i-z_j)^{1/2}
\sum_{n_N\geq \ldots \geq n_1\geq 0}
P_{\{n_N,\ldots,n_1\}}^{(-1/2)}(z_1,\ldots,z_N)
| \{n_N, \ldots n_1\} \rangle \ .
\label{phiprod}
\ee
Notice that the indices of the Jack
polynomial in the wave function are $(-\thalf)$
and $\{n\}$, which are dual to the indices
$(-2)$ and $\{n'\}$ on the Jack polynomial used
in the construction of the YHWV $|\{n\}\rangle$
in eq.~(\ref{yhw}). The Jack polynomials
are now evaluated at argument $p_n=\sum_i z_i^n$.

It will be clear to the reader that the formula
(\ref{phiprod}), which decomposes a
general spinon field product on an orthogonal
basis of YHWV's, is
a convenient starting point for the computation
of correlation functions.

Let us warm up by computing the two point function.
Assuming $|z_1|>|z_2|$, we have
\bea
\langle \phi_+(z_1) \phi^+(z_2) \rangle
&=& \langle 0| \phi_+(z_1) \phi^+(z_2) |0\rangle
\nonu
&=& z_1^{-1/2} \sum_{m,n} \, \langle m|
  P_{\{m\}}^{(-1/2)}({1 \over z_1}) P_{\{n\}}^{(-1/2)}(z_2)
  |n\rangle
\nonu
&=& \sum_n C_n^{(-{1 \over 2})} ({z_2 \over z_1})^n z_1^{-1/2}
    = (z_1-z_2)^{-1/2} \ ,
\eea
where we used that with our choice of
normalization $P_{\{n\}}^{(-1/2)}(z) = z^n$
and $j_{\{1^n\}} = C^{(-{1 \over 2})}_n$.

For a general $2N$-point function, we can avoid
using explicit expressions for the metric $j_{\{n\}}$,
and instead use the general result
\be
\sum_{\{n\}} j_{\{n'\}}
P_{\{n\}}^{(-1/2)}({x}) P_{\{n\}}^{(-1/2)}(y)
= \prod_{i=1}^{N} \prod_{j=1}^N  (1-x_iy_j)^{-1/2} \ .
\ee
This immediately leads to the following result
for the $2N$-point spinon correlator
\bea
\lefteqn{\langle \phi_+(w_1) \ldots \phi_+(w_N)
   \phi^+(z_1) \ldots \phi^+(z_N) \rangle}
\nonu
&& = \prod_{i=1}^N w_i^{-1/2}
     \prod_{i>j} ({1 \over w_i} - {1 \over w_j})^{1/2}
     \prod_{i<j} (z_i-z_j)^{1/2}
\nonu
&&\qquad \times
     \sum_{\{m\},\{n\}}
     \langle \{m\} |
     P_{\{m\}}^{(-1/2)}({1 \over w_1},\ldots,{1 \over w_N})
     P_{\{n\}}^{(-1/2)}(z_1, \ldots, z_N)
     | \{n\}\rangle
\nonu
&& = \prod_{i=1}^N w_i^{-N/2}
     \prod_{i<j} (w_i-w_j)^{1/2}
     \prod_{i<j} (z_i-z_j)^{1/2}
\nonu
&&\qquad \times
     \sum_{\{n\}} j_{\{n'\}}
     P_{\{n\}}^{(-1/2)}({1 \over w_1},\ldots,{1 \over w_N})
     P_{\{n\}}^{(-1/2)}(z_1, \ldots, z_N)
\nonu
&& = \prod_{i=1}^N w_i^{-N/2}
     \prod_{i<j} (w_i-w_j)^{1/2}
     \prod_{i<j} (z_i-z_j)^{1/2}
     \prod_{i,j} (1-{z_i \over w_j})^{-1/2}
\nonu
&& = \left( {\prod_{i<j} (w_i-w_j)
             \prod_{i<j} (z_i-z_j) \over
             \prod_{i,j} (w_j-z_i) } \right)^{1/2} \ .
\eea
One may check that this result is consistent with
the triplet channel 4-point function $G_2(x)$
that we computed in section~3. Once again, there
is no novelty in the final expressions, which
are Gaussian and can be recovered by using a
free boson formalism. Our point here has been to
illustrate some techniques which we expect to be
valuable in much more general situations.

\newsubsection{Characters}

An interesting observation, which we worked out
in our paper \cite{bls}, is that the spinon formulation
of the $SU(2)_1$ WZW model directly leads to novel
ways to write the characters of the Virasoro and
affine modules in this theory. Without repeating
the derivation here, we shall here briefly summarize
the results of \cite{bls}.

The Virasoro character for the module characterized
the conformal dimension $L_0=j^2$ (where $j=0,\thalf,1,\ldots$
can be related to an $SU(2)$ spin) is found to be

\hfill \parbox{14.5cm}{
\bea
\chi^{\rm Vir}_{j^2}(q) &=& q^{-j/2} \, \sum^*_{m_1,m_2,\ldots}
q^{\half(m_1^2+m_2^2+\ldots-m_1m_2-m_2m_3-\ldots)}
\nonu
&& \times {1 \over (q)_{m_1}} \prod_{a\geq 2}
\left[ \begin{array}{c} \half(m_{a-1}+m_{a+1}+\delta_{a,2j+1})
       \\ m_a \end{array} \right]_q \ .
\label{virj}
\eea}

\noindent In these formulas the $m_i$ are non-negative
integers; the $*$ on the summation symbol
indicates that $m_{2j},m_{2j-2},\ldots$ are to be odd and the
other $m_i$ even. The $q$-deformed binomial
coefficients that feature in this formula are
defined as
\be
\left[ \begin{array}{c} a \\ b
       \end{array} \right]_q
=
{(q)_a \over (q)_{a-b}(q)_b}, \quad {\rm for}
\ \ a \geq b \ ,
\ee
(and zero otherwise), where
\be
(q)_a = \prod_{n=1}^a (1-q^n)
\ee
with $(q)_0=1$ and $(q)_{-a}=0$. In the character
formula (\ref{virj}), the number $m_1$ is precisely
the number of spinons in a state that contributes to
the character. When expanded in a $q$-series, the
character formula (\ref{virj}) by construction agrees with
the formula (\ref{viraj}) that we wrote earlier.

The formula (\ref{virj}) is a limiting case of a
corresponding formula for Virasoro minimal models, which was
first conjectured in \cite{kedem} and later proven in
\cite{ber}.

{}From the basis of states which we gave at the end
of section~5, formula (\ref{plusmin}), one easily
derives the following expressions for the
(level-1) affine characters in the theory
\be
\chi^{A^{(1)}_1}_{j=0}(q) =
 \sum_{N^+ + N^- \; {\rm even}}
 {q^{(N^+ + N^-)^2/4} \over (q)_{N^+} (q)_{N^-}} \ ,
\qquad \qquad
\chi^{A^{(1)}_1}_{j={1 \over 2}} =
 \sum_{N^+ + N^- \; {\rm odd}}
 {q^{(N^+ + N^-)^2/4} \over (q)_{N^+} (q)_{N^-}} \ .
\ee
These formulas were first written in \cite{ezer} and
they were related to the spinon picture in \cite{bps}
(see also \cite{FS} for closely related results).

\vspace{6mm}

In summary, we have in this section indicated
a number of ways in which the spinon formulation
and Yangian symmetry of the $SU(2)_1$ CFT can be used.
We expect that many more applications are to
be found, in particular in the context of the
Thermodynamic Bethe Ansatz and of integrable
perturbations of CFT. We will report on some such
developments elsewhere \cite{tbp}.

\vspace{6mm}

{\bf Acknowledgements:}\ A.W.W.L. is an
A.P.~Sloan fellow. The research of K.S. is
supported in part by the National Science
Foundation under grant PHY90-21984.

\newpage

\appendix
\newsection{The Yangian $Y(sl_2)$}

The Yangian
$Y({\bf g})$ associated to a Lie algebra ${\bf g}$
is a Hopf algebra that is neither commutative nor
cocommutative, and as such it can be viewed as a
non-trivial example of a quantum group \cite{drin}.
Its history goes back to the general formalism of the
Quantum Inverse Scattering Method (see \cite{vladimir}
for an introduction). Indeed, the object
$Y({\bf g})$ is directly related to certain rational
solutions of the Quantum Yang Baxter Equation, the
simplest of which was first obtained by C.N.~Yang \cite{yang}.
For the purpose of these proceedings we shall consider
the case ${\bf g}=sl_N$.

We write the lowest generators of $Y(sl_N)$ as $Q_0^a$
and $Q_1^a$; higher generators can be obtained by taking
successive commutators with the generators $Q_1^a$.
The defining relations of the algebra $Y(sl_N)$ can be
written as follows \cite{drin}
\bea
{\rm (Y1)} && [Q_0^a,Q_0^b] = f^{ab}{}_c Q_0^c \ ,
\nonu
{\rm (Y2)} && [Q_0^a,Q_1^b] = f^{ab}{}_c Q_1^c \ ,
\nonu
{\rm (Y3)} && [Q_1^a,[Q_1^b,Q_0^c]] + ({\rm cyclic}\;\;{\rm  in}\ a,b,c)
     = A^{abc}{}_{def} \{ Q_0^d, Q_0^e, Q_0^f \} \ ,
\nonu
{\rm (Y4)} && [[Q_1^a,Q_1^b],[Q_0^c,Q_1^d] + [[Q_1^c,Q_1^d],[Q_0^a,Q_1^b]]
\nonu
&& \qquad
     = \left( A^{abp}{}_{qrs} f^{cd}{}_p + A^{cdp}{}_{qrs} f^{ab}{}_p \right)
       \{ Q_0^q, Q_0^r, Q_1^s \} \ ,
\label{terrific}
\eea
where $A^{abp,def} = {1 \over 4} f^{adp} f^{beq} f^{cfr} f_{pqr}$ and
the curly brackets denote a completely symmetrized product.
The $SU(N)$ structure constants $f^{abc}$ have been normalized
as
\be
   f^{abc} f^d{}_{bc} = -2N \, d^{ad} \ .
\ee
The following comultiplications may be used to define
the action of the Yangian generators on a tensor product
of states
\bea
\label{copr}
&& \Delta_{\pm}(Q_0^a) =
  Q_0^a \otimes {\bf 1} + {\bf 1} \otimes Q_0^a \ ,
\nonu
&& \Delta_\pm(Q_1^a) =
  Q_1^a \otimes {\bf 1} + {\bf 1} \otimes Q_1^a
  \pm \thalf f^a{}_{bc} Q_0^b \otimes Q_0^c \ .
\eea
The `terrific' (dixit Drinfel'd \cite{drin}) right hand
sides of the relations (Y3) and (Y4) can be derived from
the homomorphism property of these comultiplications.
For ${\bf g}=sl_2$, the cubic relation (Y3) is superfluous
and for all other algebras (Y4) follows from (Y2) and (Y3).

\end{document}